# Highly efficient optical transition between excited states in wide InGaN quantum wells

G. Muziol,[1,*] H. Turski,[1] M. Siekacz,[1] K. Szkudlarek[1], L. Janicki,[2] S. Zolud,[2] R. Kudrawiec,[2] T. Suski,[1] and C. Skierbiszewski,[1,3]

[1]*Institute of High Pressure Physics PAS, Sokolowska 29/37, 01-142 Warsaw, Poland*

[2]*Faculty of Fundamental Problems of Technology, Wroclaw University of Science and Technology, Wybrzeze Wyspianskiego 27, 50-370 Wroclaw, Poland*

[3]*TopGaN Ltd, Sokolowska 29/37, 01-142 Warsaw, Poland*

*E-mail: gmuziol@unipress.waw.pl

**Abstract**

There is a lack of highly efficient light emitting devices (LEDs) operating in the green spectral regime. The devices based on (In,Al)GaN show extremely high efficiencies in violet and blue colors but fall short for longer emission wavelengths due to the quantum confined Stark effect (QCSE). In this paper we present a design of the active region based on wide InGaN quantum wells (QWs) which do not suffer from QCSE and profit from an enhancement in the internal quantum efficiency (IQE). The design exploits highly efficient optical transitions between excited states. It is shown that, counterintuitively, the devices with higher InGaN composition exhibit a higher enhancement in IQE. Experimental evidence is provided showing a gradual change in the nature of the optical transition with increasing thickness of the QW. Moreover, optical gain in long wavelength LDs incorporating standard and wide QWs is investigated to show the utilization of our concept.

## Introduction

The optoelectronic devices based on III-nitrides had found many applications including general lighting and displays thanks to their high quantum efficiency [1-7]. One of the key challenges in the design of the active region of the devices are caused by the strong built-in electric fields arising both from the spontaneous and piezoelectric polarization [8-11]. Although this built-in electric field enabled the realization doping-free high electron mobility transistors [12-16] and polarization-induced doping [17] it can be detrimental in the optoelectronic devices like light emitting devices (LEDs) and laser diodes (LDs). The active region of LEDs and LDs consists of GaN/AlGaN or InGaN/GaN multi quantum well (MQW) structures. Due to the lattice mismatch between the alloys a strong piezoelectric field is present in the QWs. It causes the spatial separation of electron and hole wavefunctions leading to red-shift of the emission spectra, called the quantum confined Stark effect (QCSE), and reduction of the wavefunction overlap [10, 18-26]. The latter leads to an increase of the carrier density because of a lower probability of carrier recombination. This is unwanted because as the carrier density increases a growing part of the carriers recombine through the nonradiative Auger process causing the reduction of the quantum efficiency [27-31]. Additionally, in case of the InGaN MQWs the higher the indium content the higher the polarization field and thus the separation of carrier wavefunctions. This is the primary reason for the loss of efficiency of the III-nitride devices in the green spectral range [5, 32].

From the point of view of reducing the impact of Auger recombination it would be ideal to reduce the carrier density. A simple increase of the active region volume by increasing the number of QWs does not provide the expected efficiency enhancement because of the non-uniform hole distribution among QWs [33]. On the other hand, the increase of the QW thickness, due to the polarization-induced separation of carrier wavefunctions, decreases the transition probability severely [19, 34]. The wider the QW the lower the wavefunction overlap is. Several methods of overcoming the limitation arising due to the built-in electric field had been proposed. The two major are growth of staggered QWs [35, 36] and growth on semipolar and nonpolar crystal orientations [37-42].

There are a few reports in the literature studying the thick InGaN QWs on the polar c-plane orientation [19, 22, 34, 43, 44] and, although the calculations predict a severe decrease of the transition probability [19, 34], some of the results show an increase in the efficiency [44]. Additionally, the current state-of-the-art violet LDs incorporate wide (6.6 nm thick) InGaN QW [45]. There clearly is a discrepancy between the theoretical understanding and the experimental results.

In this paper it will be shown that, counterintuitively, the LEDs and LDs with wide InGaN QWs can have higher efficiency than with the thin QWs. The enhancement in the efficiency is caused by two factors: (i) screening of the built-in electric field by carriers occupying the ground states and (ii) emergence of exited states with highly efficient optical transition. Together with the lower carrier density in the wide QWs it significantly reduces the non-radiative Auger recombination making the wide InGaN QWs a good candidate for efficient sources of light in the green spectrum. Additionally, it will be presented, contrary to the common understanding, that the higher the composition of the QW the higher this increase in the efficiency is.

The scheme of this paper is as following: (1) theoretical calculations will be presented to study the influence of thickness on the optical transitions inside the QW and explicate the idea behind the increased efficiency, (2) photoluminescence spectra of QWs with various thicknesses showing the experimental evidence of emergence of excited states will be shown, (3) the influence of indium composition on internal quantum efficiency (IQE) will be studied theoretically and (4) a comparison of optical gain of blue and cyan LDs with thin and wide QW will be shown to validate the predicted change in the efficiency with the InGaN composition.

## Appearance of efficient transitions through excited states in wide InGaN QWs

The intriguing effect which triggered our interest in wide InGaN QWs was a peculiar dependence of photoluminescence (PL) intensity on QW thickness. Fig 1(a) presents the PL intensities for $In_{0.17}Ga_{0.83}N$ single QW samples grown by plasma assisted molecular beam epitaxy (PAMBE) with the thickness varied from $d_{QW}$=1.2 to 25 nm. The structure of the single QW is shown as an inset to Fig 1(a). Details on the PAMBE growth technique can be found elsewhere [46, 47]. The indium content in InGaN quantum barriers (QBs) was chosen to be 8% as the LDs presented in the latter part of this paper also have $In_{0.08}Ga_{0.92}N$ QBs. The presence of indium in QBs slightly lowers the built-in electric field but does not change the qualitative trends. The In content in the QWs was corroborated by X-ray diffraction and is constant among the samples. The QWs were grown using high nitrogen flux RF plasma source which, as we demonstrated earlier, leads to high quality InGaN growth in PAMBE [48-50]. Fig 1(b) shows the transmission electron microscopy picture of the 15.6 nm QW indicating that both QW interfaces are sharp and the overall good structural quality. It is important as it shows that the samples of study can be treated as uniform and high quality QWs and not for example quantum dots. Depending on the excitation power used of the 325 nm He-Cd laser the measured intensity was observed to decrease or increase with the increasing QW thickness. According to the QCSE as the QW thickness is increased the wavefunction overlap and thus intensity of PL should drop. This is observed if a relatively small 0.4 W/cm$^2$ excitation power is used. However, when the excitation power is increased to 60 W/cm$^2$ an opposite trend is seen which comes from screening of the built-in electric field by generated carriers and emerging of an efficient transition through excited states.

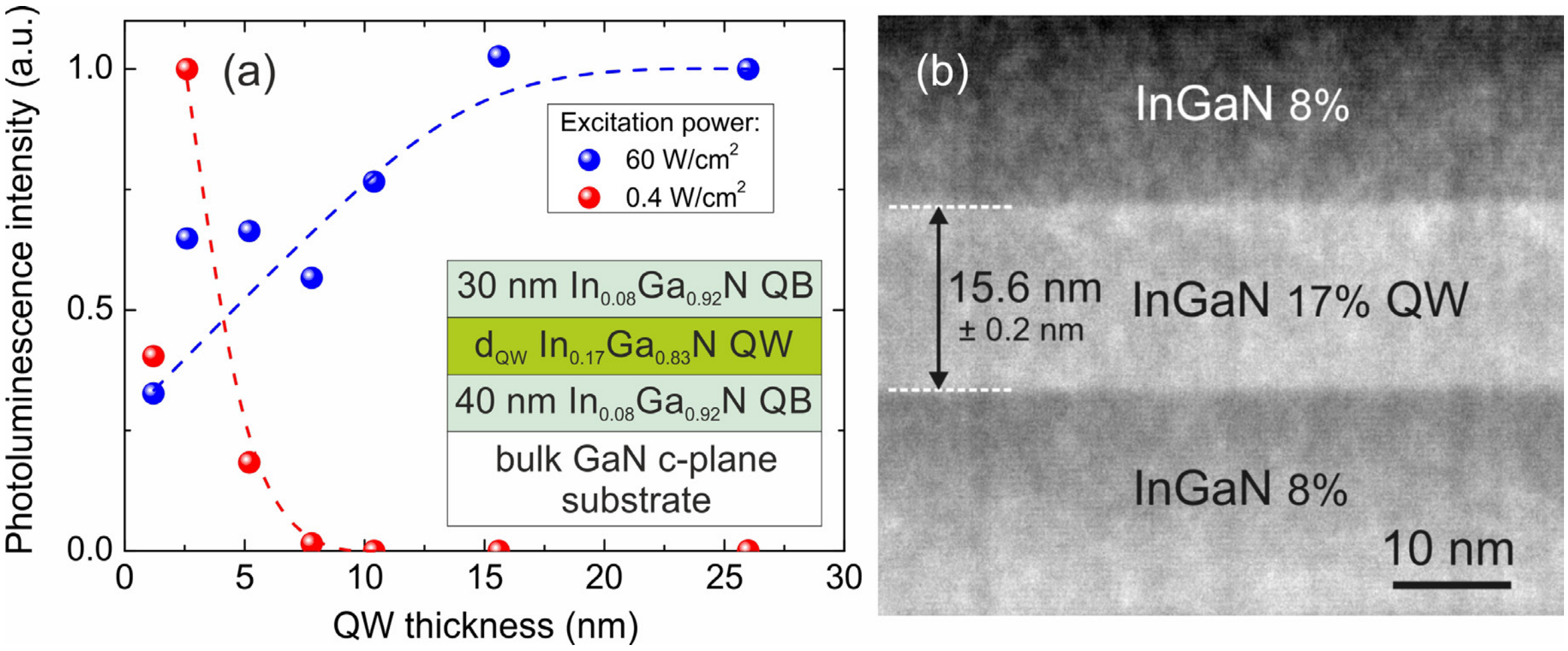

*Fig 1. (a) Dependence of photoluminescence intensity on QW thickness for two excitation power densities. The dashed lines are guides to the eye and present two contradictory trends. (b) TEM image of the 15.6 nm QW showing sharp bottom and top interfaces.*

To gain insight into the mechanisms responsible for the observed changes in the trends the SiLENSe 5.4 package has been used for calculation of the potential shape and wavefunction distribution at various excitations [51]. Although SiLENSe package requires an LED structure as input it can be used to a certain extent to mimic the behavior of QW samples under optical excitation. The comparison of the carrier wavefunction overlaps between thin and wide QWs has been shown in Fig 2(a). The ground transition *<e1h1>* of thin QWs has an initial overlap of 0.31 at j=0 A/cm$^2$ which increases with current density due to screening of the built-in polarization fields. In the case of wide QW, as it had been shown earlier, the transition between ground states is significantly lower than the thin counterpart due to separation of electron and hole wavefunctions driven by QCSE. However, there is a significant transition path including the excited states *e2* and *h2*. The wavefunction overlap of the *<e2h2>* transition, although initially nearly equal to 0, peaks to a 0.63 at j=300 A/cm$^2$ which is higher than the overlap of the ground transition of the thin QW. The crucial factor in observing this high efficiency transition is that the excited states in the wide QW get populated with carriers. This will be covered later.

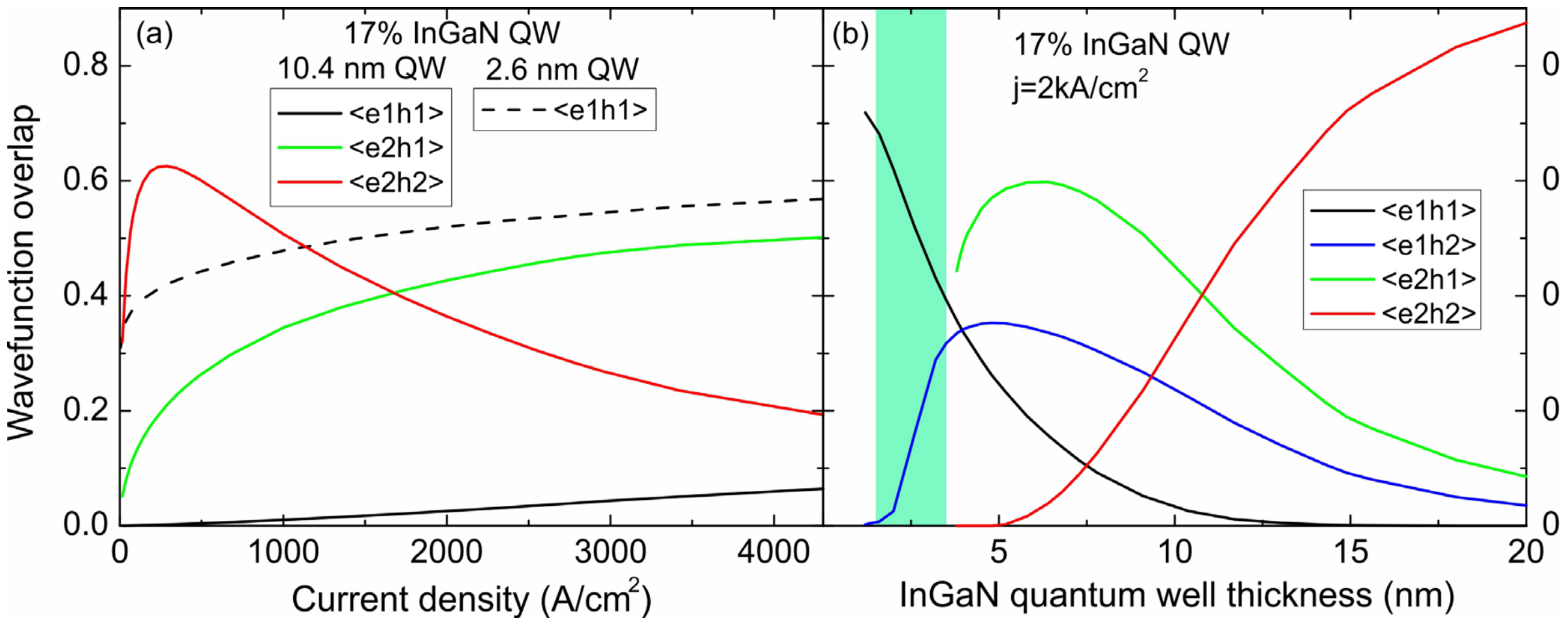

*Figure 2. (a) Wavefunction overlaps within QWs of two thicknesses: a thin 2.6 nm and a wide 10.4 nm as a function of current density. Only three types of transitions <e1h1>, <e2h1> and <e2h2> are shown to clearly present the idea behind the high efficiency of wide QWs. (b) The full evolution of the wavefunction overlaps of four transitions as a function of well thickness. The region marked with a green rectangle shows the commonly used QW thicknesses. The emerging of a high efficiency transition <e2h2> is observed for large thickness of the QW.*

The full evolution of the wavefunction overlaps with the QW thickness is shown in Fig 2(b) for a fixed current density j=2kA/cm$^2$. This value is chosen on purpose to show the QW band profile close to lasing thresfhold. The standard operating conditions for LEDs are much lower (j=10A/cm$^2$) due to the efficiency droop. However, the efficiency droop of LEDs with wide QWs, as will be shown later, is lower. Therefore they can be operated with success at higher current densities. The green-colored region, ranging from 1.5 to 3.5 nm, marks the thicknesses of the QWs commonly used in LEDs and LDs. In the thin QW regime, even up to a thickness 4.1 nm, the overlap between the ground states <*e1h1*> decreases with the QW thickness leading to a higher carrier concentration for the same radiation rate. However, above 4.1 nm an excited state in the conduction band *e2* appears with an initially high overlap with the valence band ground state *h1*. The high recombination probability of the <*e2h1*> transition is counterintuitive as it would violate the selection rules of a rectangular QW. However, the selection rules do not apply to the case of InGaN QWs with high built-in electric field which breaks the symmetry of the potential. As the thickness is increased the overlap between the excited states <*e2h2*> rises whereas other overlaps drop. The value of the <*e2h2*> overlap of wide QW becomes higher than the value of <*e1h1*> overlap of thin QW. At the same time the carrier density of the wide quantum well can be much lower as compared to the thin counterpart. This leads to a reduction of the non-radiative Auger recombination and a much more efficient radiative transition.

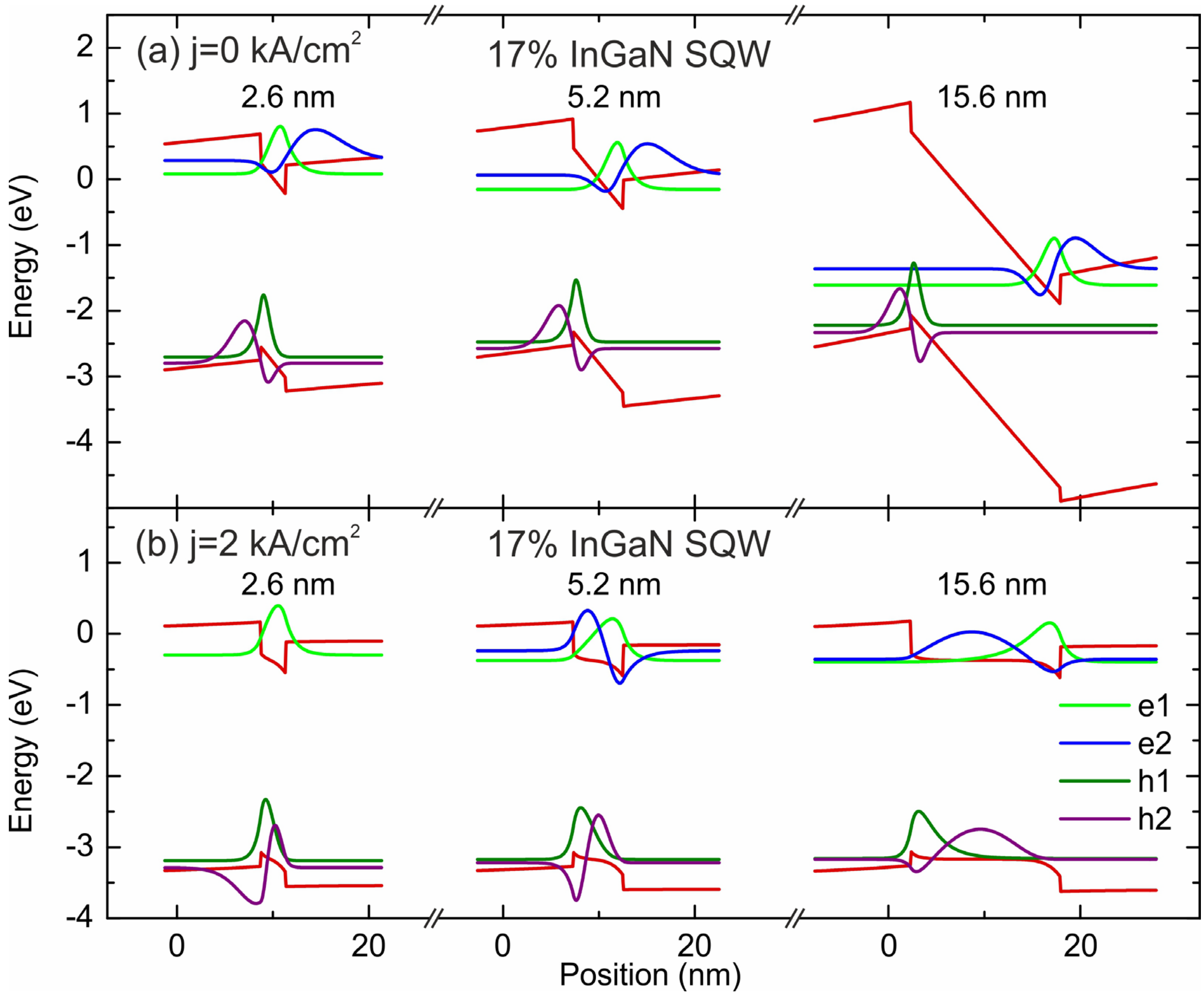

*Figure 3. Band profiles and carrier wavefunctions of 17% InGaN QWs in the (a) no injection and (b) high injection regimes. The full screening of polarization field and high wavefunction overlap of excited states is observed only for the wide QW. The e2 state does not form in case of the 2.6 nm thick QW at high injection due to the insufficient barrier height.*

Three regimes can be distinguished from Fig 2(b): thin QWs up to 4.1 nm where only *<e1h1>* and *<e2h1>* transitions can be observed, intermediate QWs up to 10 nm with all four transitions, and wide QWs above 10 nm where the *<e2h2>* transition exceeds the others. To understand the interplay between these regimes the band profiles and carrier wavefunctions are presented in Figures 3(a) and 3(b) for current densities of j=0 A/cm$^2$ and j=2×10$^3$ A/cm$^2$, respectively. QWs of three thicknesses: 2.6 nm, 5.2 nm and 15.6 nm, have been presented to explain the behavior of the three regimes presented in Fig 2(b). First, the case in which no current flow is present will be discussed. As can be seen in Figure 3(a) a strong built-in field of ~1.7 MV/cm is present in the QWs which causes the separation of the electron and hole wavefunctions and the QCSE. The wavefunction overlaps between the electron and hole ground states are 0.19, 0.0006 and 10$^{-24}$ for the QWs with thicknesses of 2.6, 5.2 and 15.6 nm, respectively. In the second case presented in Figure 3(b) in which j=2×10$^3$ A/cm$^2$ the wavefunction overlaps between the ground states are equal to 0.52, 0.23 and 0.0007. Although the overlaps are slightly higher due to the screening of the built-in fields still a dramatic decrease can be observed with increase of the QW thickness. This decrease is the cause of the common opinion that the thicker the InGaN QW the lower the efficiency of any device build based on it. On the other hand, the excited states show a remarkably high wavefunction overlaps. The nature of the built-in electric field is responsible for this discrepancy and will be discussed below.

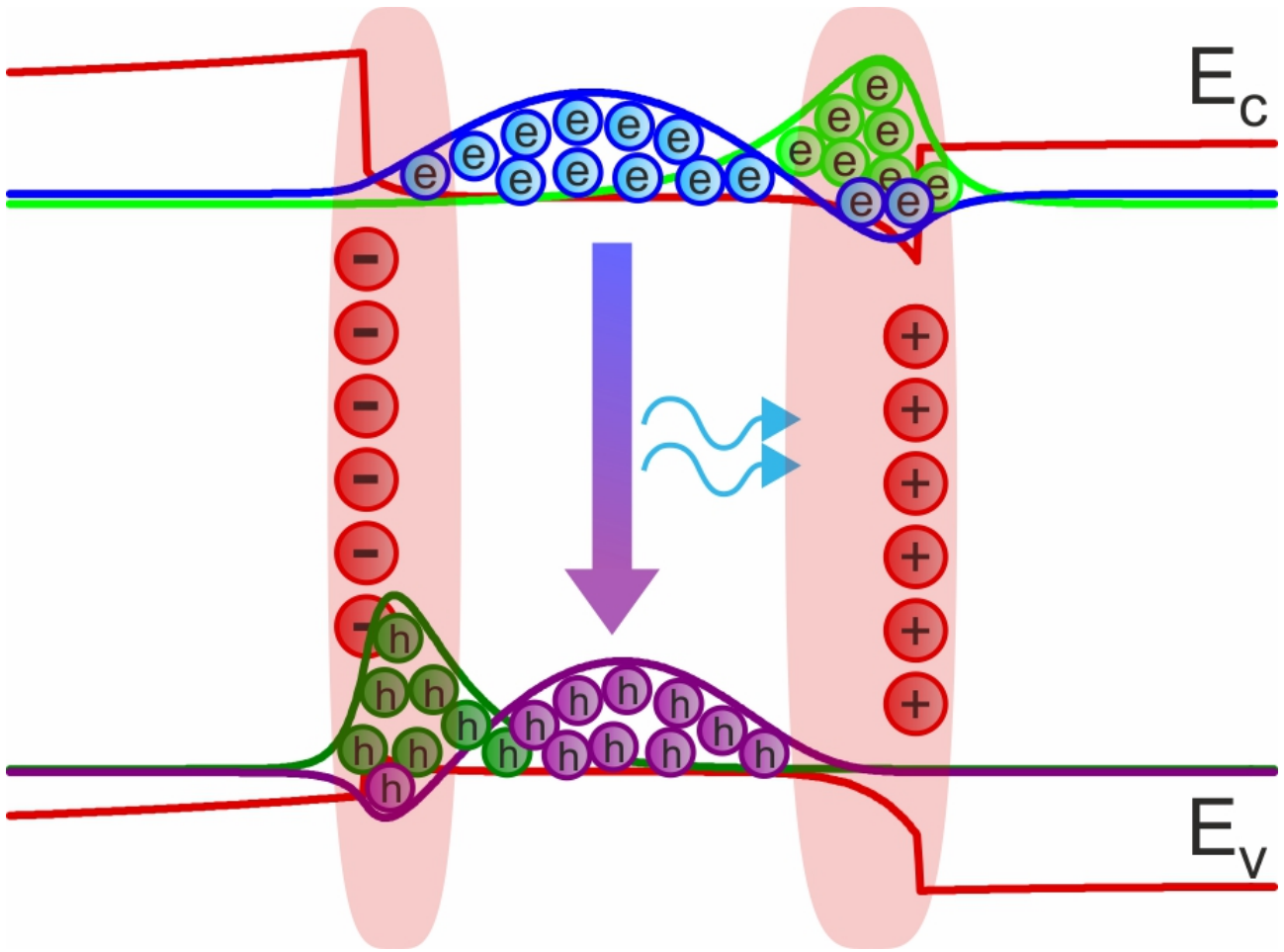

*Figure 4. A schematic of the band profile and carrier distribution in the 15.6 nm wide QW at j=2 kA/cm2. The carriers occupying the e1 and h1 do not recombine due to almost zero overlap between their wavefunctions and build up their concentrations. This lead to screening of the piezoelectric sheet charges, marked with red circles. The red shaded area depicts the part of the QW in which there is still some residual electric field. The carriers on e2 and h2 states are involved in the radiative transition. The distribution along the whole QW thickness of carriers on e2 and h2 states ensures the low carrier density and reduced recombination through the Auger process.*

The built-in field is caused by appearance of fixed electric charges at the QW interfaces. For the 17% InGaN QW used in this example the polarization-induced sheet charges of $1\times10^{12}$ cm$^{-2}$ form at the QW interfaces as indicated in Fig 4. To screen the polarization charges a matching number of electrons and holes has to be introduced. However, the electron and hole wavefunctions have a nonzero spatial dimension. Therefore, even if a matching number of electrons and holes is introduced, the remains of the polarization field are present locally. This causes the notable difference between the thin (2.6 nm) and thick (15.6 nm) QWs. The distance at which the influence of fixed charges is present in case of the thin QW is comparable to its thickness. The screening of the fixed electric charges changes the values of the wavefunction overlaps but the qualitative “triangular-like” appearance of the band profile of QW is similar (compare Fig 3 (a) and Fig 3(b)). On the other hand, in case of the wide QW the traces of the screened built-in field are present only in the bending of the conduction and valence bands at the QW interfaces (see shaded area in Fig 4). In the middle of the wide QW no field is present and, in this region, its qualitative appearance is as of a square QW. Therefore, the ground transition <*e1h1*> is weak but the transitions between excited states <*e2h2*> is strong almost as in the case of a rectangular QW without piezoelectric fields.

The carrier density required to match and fully screen the polarization charges is $5 \times 10^{19}$ $cm^{-3}$. To benefit from the high-overlap excited states transition <*e2h2*> can be achieved only if the polarization field is screened. The required carrier density seems to be extremely high as it is comparable to the one observed in III-nitride LDs at threshold [52]. However, considering the low transition probability before the screening of piezoelectric field this high carrier density can be achieved easily as long as no other carrier loss mechanisms such as recombination via defects or thermal escape play a significant role. The initial carrier accumulation behavior of a thick QW can be thought of as a capacitor. Afterwards, a strong <*e2h2*> transition emerges and the build-up of carriers is suppressed.

Below the mechanism of filling of the excited states with carriers will be discussed. At no carrier injection, the case shown in Fig 3(a), the energetic distances between the ground and excited states are slightly increasing with QW thickness and are equal for the 2.6 nm thick QW to 200 and 90 meV for electrons and holes, respectively. However, after the screening of the polarization field the difference in the energy level between the ground and excited states decrease with the increase of QW thickness as expected for a square-like potential shape (compare Fig 3(b)). In the case of the thin 2.6 nm QW the *e2* excited state does not even form because of its energy level being higher than the quantum barrier while the difference in the energy levels of *h1* and *h2* is still 90 meV whereas in the case of 15.6 nm thick QW the energy level differences are equal to 39 and 13 meV for electrons and holes, respectively. Taking into account the high carrier density, required to screen the piezoelectric field, and the density of states of *e1* and *h1* it can be clearly shown that the carriers will start to occupy the *e2* and *h2* states. Assuming a bulk density of states the quasi-Fermi levels of the conduction and valence band for a carrier density of $5 \times 10^{19}$ $cm^{-3}$ would be 230 and 20 meV, respectively. This is far above the energy level difference between the ground and excited states, therefore the excited states will be occupied as soon as the carrier density reaches value required to screen the polarization charges and screen the built-in electric field. This is feasible as the carriers occupying the ground states do not recombine because of the low transition probability. Eventually, carriers will start to occupy the excited states.

## Photoluminescence experiments

To verify the predictions of the model presented above a set of 17% InGaN QWs with thicknesses ranging from 1.2 to 25 nm had been prepared by PAMBE. In this section only three

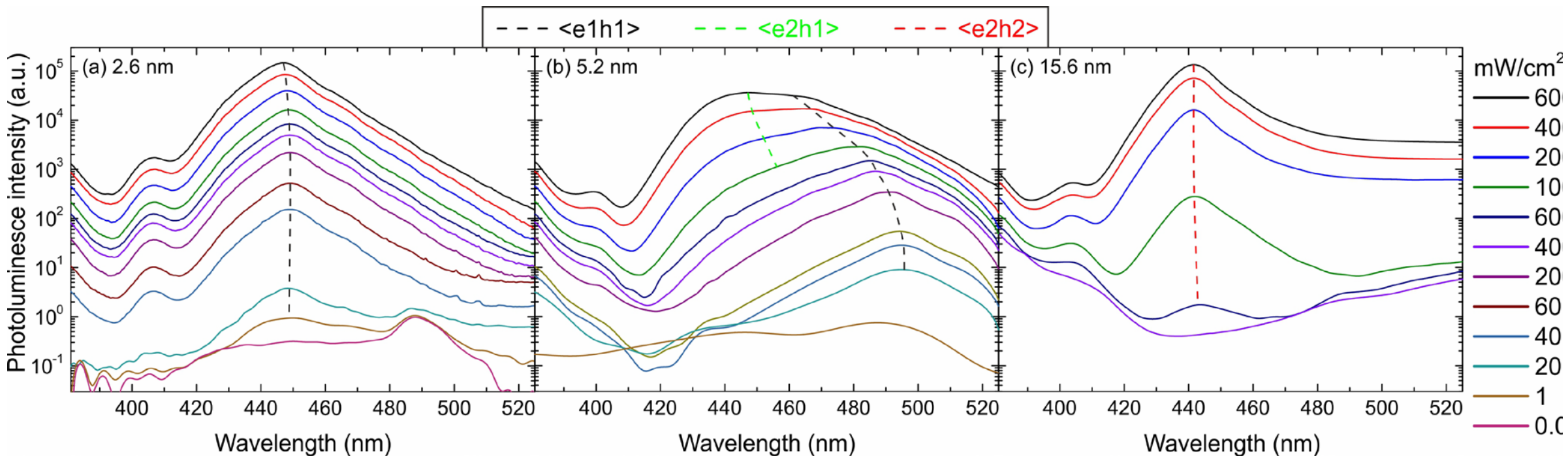

*Figure 5. Power dependent photoluminescence for QWs with three thicknesses: (a) 2.6 nm, (b) 5.2 nm and (c) 15.6 nm. In case of the thin 2.6 nm QW only the <e1h1> transition is observed, whereas in case of the intermediate 5.2 nm QW a second transition interpreted as the <e2h1> appears and is depicted with a green dashed line. The thick QW starts to emit light at a relatively high excitation through the <e2h2> transition.*

QWs with thicknesses representing the three regimes observed in Fig 2(b) are shown. However, the behavior shown below is persistent in all of the studied samples. The PL data of all of the studied samples is provided in the Supplementary Figure 1. The power dependent PL of the three samples are shown in Fig 5(a), 5(b) and 5(c). The excitation power regime is extremely low ranging from 40 μW/cm$^2$ up to 6 W/cm$^2$ and has been chosen to show the behavior at the lowest carrier densities at which the radiative transitions were detectable. The thin 2.6 nm QW, shown in Fig 5(a), starts to emit light at $\lambda$=450nm and slightly shifts to $\lambda$=447nm due to screening of the polarization field and the band-filling effect. At all excitation powers only one transition *<e1h1>* is observed as expected when analyzing Fig 2(b). The peak observed at 490 nm comes from the residues in the He-Cd laser used to excite the samples and can be observed at extremely low light intensities. The peaks at $\lambda$=400-405 nm observed also in all of the other samples come from the $In_{0.08}Ga_{0.92}N$ layers. In the case of the intermediate 5.2 nm QW the emission starts at a wavelength of $\lambda$=495 nm. The large difference between the emission wavelength of 2.6 and 5.2 nm thick QWs is due to the QCSE which red-shifts the emission wavelength of the latter. However, with the increase of excitation power the emission blue-shifts due to screening of the built-in electric fields. Additionally, starting from the excitation power of 1 W/cm$^2$, emerging of a second higher energy transition can be seen indicated with a green dashed line in Fig 5(b). Its intensity is growing extremely fast with power and at the maximum power used 6 W/cm$^2$ is even higher than the ground transition. This is expected as the wavefunction overlap of the transitions including excited states is much higher than the *<e1h1>* transition. The energy difference between the two transitions is 100 meV. This value together with the fact that the *<e2h1>* has the highest wavefunction overlap suggest that it is the observed transitions. However, the calculated energy difference between *h1* and *h2* levels is small which does not allow to unambiguously rule out the *<e2h2>* transition. The luminescence of the thickest, 15.6 nm, QW starts at λ=444 nm and shifts to λ=442 nm at higher excitation power. This peak is attributed to be the *<e2h2>* transition based on the fact that it is the only allowed transition as seen in Fig 2(b). The important issue is the power density at which the emission from the QW starts to be observed. The excitation density of 600 mW/cm$^2$ is much higher than in the case of thinner QWs in which the *<e1h1>* transition could be observed. Our interpretation is that until the polarization field is screened the radiative recombination is suppressed due to nearly zero wavefunction overlap. Due to thermal escape and non-radiative recombination via defects a certain level of excitation is needed for the carrier density to build up and screen the polarization charges. Afterwards, the profile of the thick QW become square-like, as in Fig 3(b), and the highly efficient *<e2h2>* starts to dominate.

The above experiment proves the predictions of the model of efficient recombination via excited states in wide InGaN quantum well.

## Influence of composition on wavefunction overlap – towards efficient green emitters

In this section the behavior of the transitions including the excited states for high indium content QWs will be discussed. It will be shown that the efficiency drop observed in the green spectral regime can be compensated by the use of wide InGaN QWs. This is contrary to the common understanding as the higher indium content generates a higher misfit between GaN substrate and InGaN QW and thus enhances the piezoelectric fields. The increased piezoelectric field causes a growing separation of electron and hole wavefunction and decreases their overlap. Fig 6(a) presents the calculated dependence of wavefunction overlap between the *<e1h1>* on the composition of a thin 2.6 nm QW. The commonly observed drop of efficiency is attributed mainly to the decrease of wavefunction overlap with indium content [32].

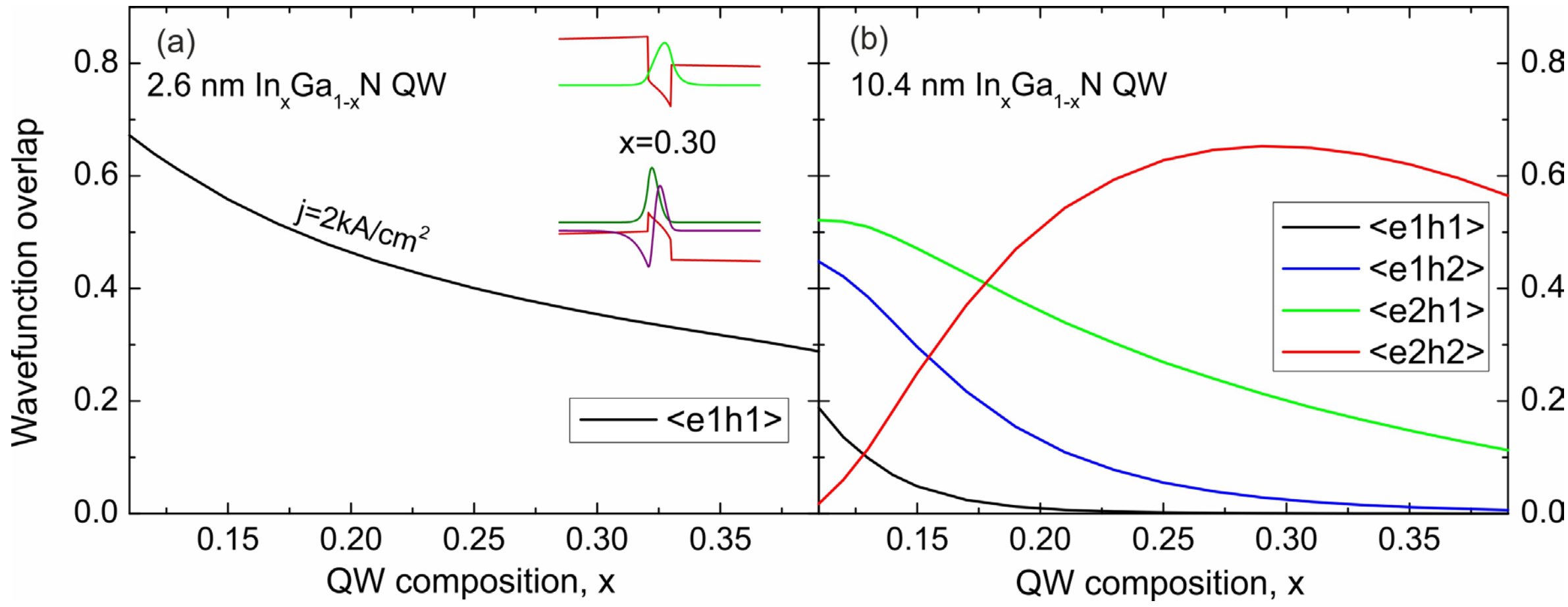

*Figure 6. The dependence of wavefunction overlap on the composition of the: (a) thin 2.6 nm and (b) wide 10.4 nm QW. The <e1h1> transition probability drops with indium content in both cases whereas the <e2h2> highly increases for the wide QW. Inset to Fig 5(a) presents the band profile and carrier wavefunctions of a 30% InGaN QW. The energetic distance between the h1 and h2 levels is large enough to prevent occupation of the h2 level.*

Naturally, one would expect to observe a similar behavior if the thickness of the QW is changed. However, as shown in Fig 6(b), the transitions including excited states behave differently. Although the probability of the transition between ground states <*e1h1*> decreases, the <*e2h2*> overlap increases with thickness. This is surprising and counterintuitive. The <*e2h2*> overlap of the wide 10.4 nm QW reaches a maximum value of 0.65 at 30% indium content. This value is strikingly high and together with the low carrier density characteristic for a wide QW makes it a good candidate for efficient green light sources.

To compare the efficiency of the thin and wide QWs the following calculation has been carried out. The IQE is given by:

$$IQE = \frac{\Gamma_{eh} B n^2}{\Gamma_{eh} A n + \Gamma_{eh} B n^2 + \Gamma_{eh} C n^3} \quad (1),$$

where $\Gamma_{eh}$ is the overlap between electron and hole wavefunctions, $n$ is the carrier density and $A$, $B$ and $C$ are the Shockley-Read-Hall, radiative and Auger recombination coefficient, respectively. The IQEs for $In_{0.17}Ga_{0.83}N$ (blue color regime) and $In_{0.30}Ga_{0.70}N$ (green color regime) has been calculated for the ABC parameters reported in [53] and plotted in Fig 7(a). As can be seen from Eq. (1) the $\Gamma_{eh}$ term cancels out and thus the IQE is independent on the overlap nor on the thickness of the QW. To understand how the overlap and QW thickness influence the efficiency it is important to look at the relationship between carrier density and current densities:

$$j = q d_{QW} (\Gamma_{eh} A n + \Gamma_{eh} B n^2 + \Gamma_{eh} C n^3) \quad (2),$$

where $q$ is the elementary charge and $d_{QW}$ is the QW thickness. The higher the thickness and the higher the overlap the lower the actual carrier density is as can be derived from Eq (2). The reduction in carrier density cannot be realized by an increase of the number of QWs because the carrier recombination takes place only at the last QW due the high difference in electron and hole mobilities [33, 44].

Fig 7(b) presents the calculated dependence of carrier density on current density in 2.6 nm and 10.4 nm thick InGaN QWs with two In compositions. The calculated change of wavefunction overlap with current density has been taken into account. For each QW only the highest value wavefunction overlap has been taken into account meaning that for the 2.6 nm $In_{0.17}Ga_{0.83}N$ the carrier density is calculated assuming only the <*e1h1*> transition while for the 10.4 nm $In_{0.17}Ga_{0.83}N$ initially the <*e2h2*> overlap is taken and is changed to the <*e2h1*> transition above j= 1.6 kA/cm$^2$ (please compare Fig 2(a)). The transitions for 2.6 and 10.4 nm

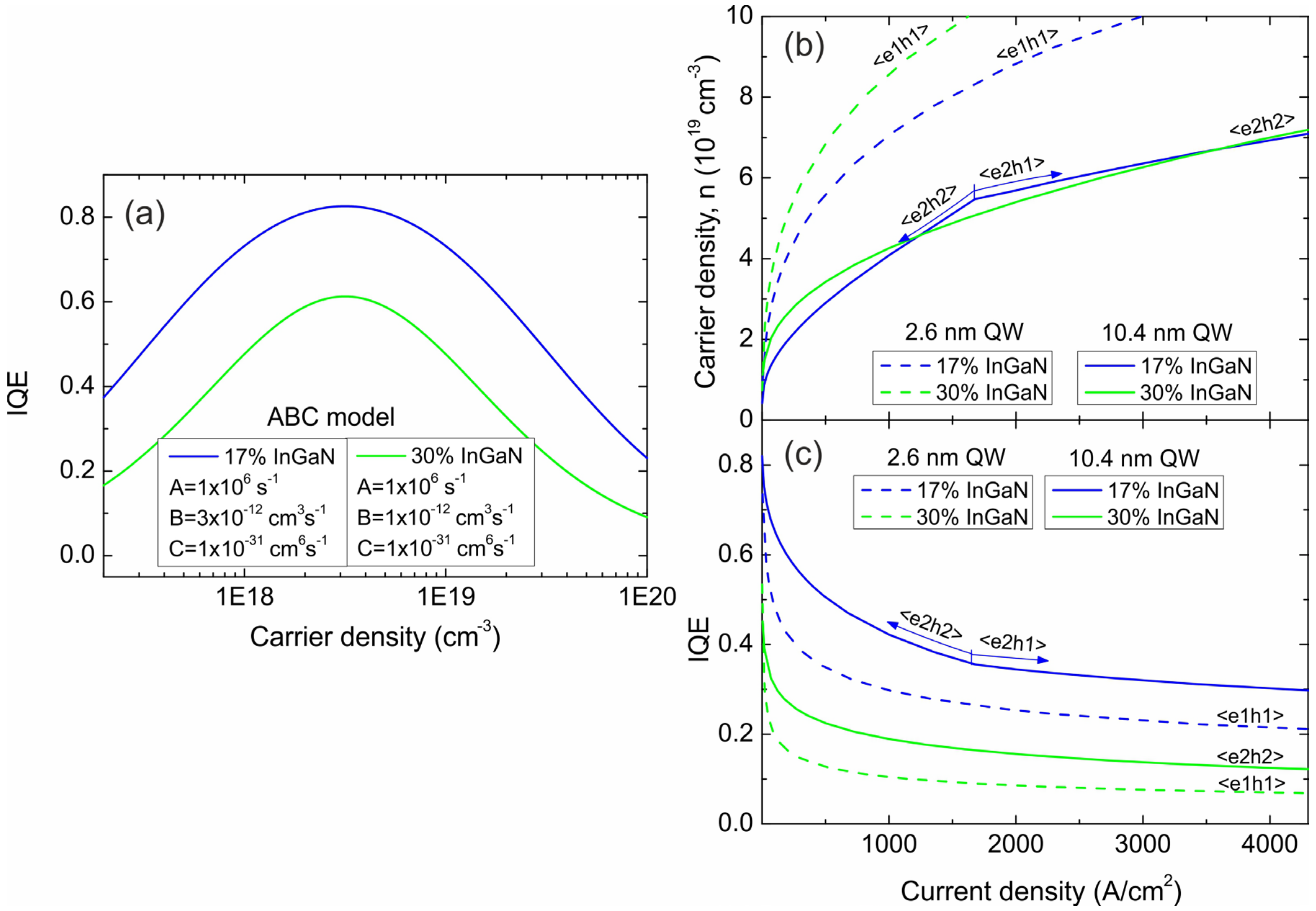

*Figure 7. (a) Dependence of the internal quantum efficiency on carrier density for $In_{0.17}Ga_{0.83}N$ and $In_{0.30}Ga_{0.70}N$ QWs. It is important to stress that it is independent on the thickness of the QW. (b) Relation between carrier density and current density for 4 kinds of QWs calculated taking into account the change in active region volume and wavefunction overlap. (c) Resultant dependence of internal quantum efficiency on current density. Only one transition with the highest wavefunction overlap is taken into account. In case of 10.4 nm $In_{0.17}Ga_{0.83}N$ QW the dominant transition changes from <e2h2> to <e2h1> which is indicated in (b) and (c).*

$In_{0.30}Ga_{0.70}N$ are *<e1h1>* and *<e2h2>* in the whole current density range, respectively. This approximation is sufficient and does not change qualitatively the results presented in this paragraph. The difference in carrier densities at a given current density causes changes to the efficiency. The calculated dependence of IQE on current density is presented in Fig 7(c). The results show a well-known decrease of the IQE with the current density for all QWs (commonly referred to as “droop”) and a decrease of the IQE with In content of the QW (commonly referred to as “the green gap”). Additionally, it can be seen that the wide QWs provide a much higher IQE in the high current regime for both cases of blue ($In_{0.17}Ga_{0.83}N$) and green ($In_{0.30}Ga_{0.70}N$) emitting LEDs. The reason is the decreased carrier density and thus the part of carriers lost to recombination through the nonradiative Auger process. The increase in IQE coming due to the use of 10.4 nm QW at j=1 kA/cm$^2$ is 40% and 70% for the $In_{0.17}Ga_{0.83}N$ and the $In_{0.30}Ga_{0.70}N$ QW, respectively. It is surprising and counterintuitive that the increase in IQE is higher for the higher indium content QW.

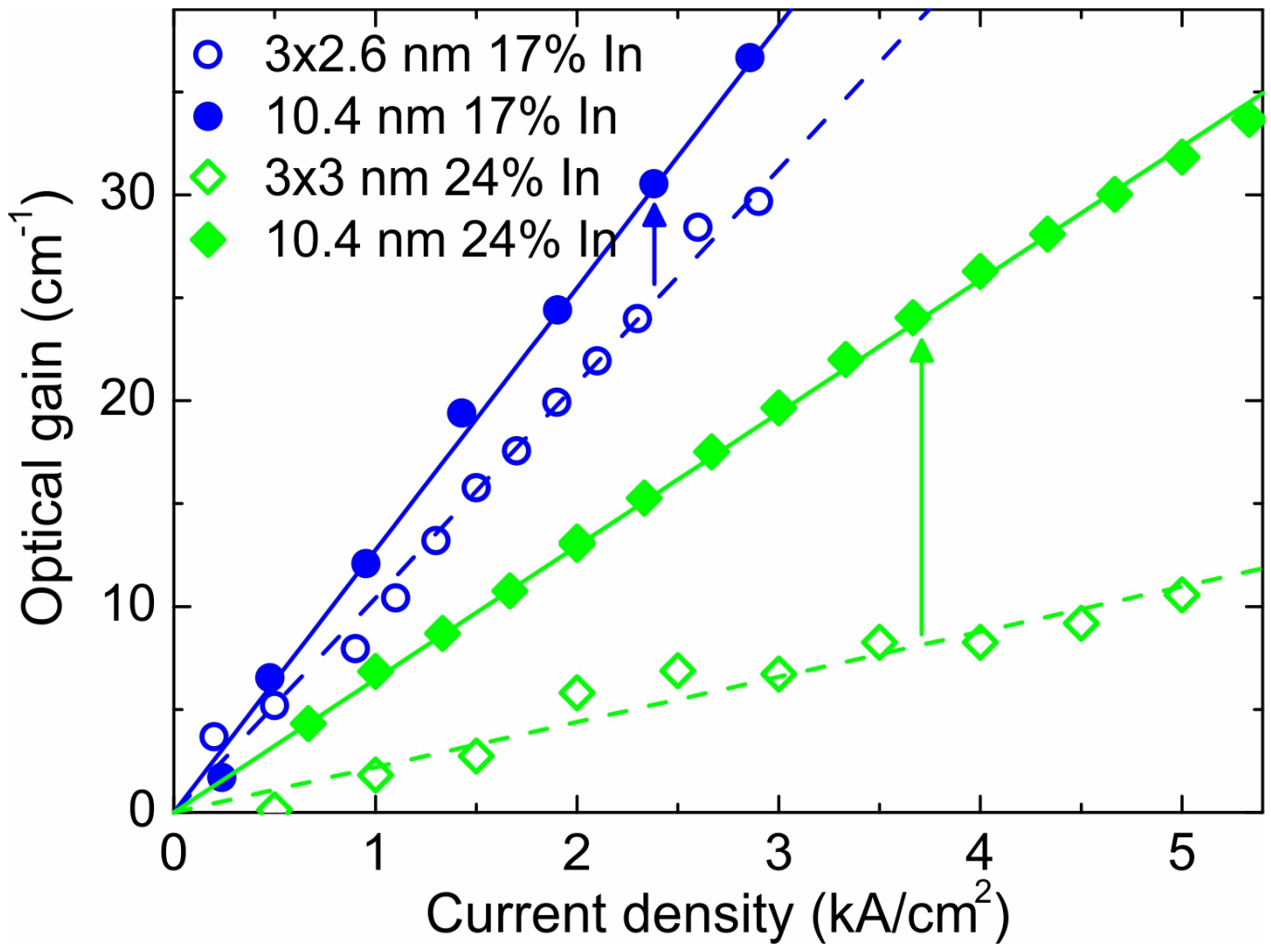

*Figure 8. (a) Optical gain of 4 LDs with different QWs. The MQW LDs consist of three 2.6 nm thick QWs separated by 8 nm thick $In_{0.08}Ga_{0.92}N$ barriers. The SQW LDs have a single 10.4 nm wide QW. Solid and dashed lines are used to extract the differential gain. The arrows indicate the increment achieved due to the use of the wide QW.*

## Optical gain of wide InGaN QWs

To verify the theoretical predictions of high efficiency long wavelength emitters based on wide QWs a set of four LDs had been manufactured. The measurement of optical gain is used to show the change in the efficiency and that the wide QWs can be used in devices requiring high carrier concentration. The optical gain in InGaN LDs has been widely studied and it was shown that it suffers from the green gap problem just as LEDs do [54-57]. There are two major causes of the decrease of optical gain for long wavelength LDs: (i) droop of quantum efficiency and (ii) lower optical confinement factor (Γ). The decrease of Γ is due to lower refractive index contrast between the alloys as the operating wavelength increases [58]. It cannot be easily increased by increasing the QW number because of the, mentioned above, non-uniform carrier distribution among QWs. However, the increase of the QW thickness should allow for the increase in Γ as long as there exists the proposed efficient transition through excited states.

The LD structures and calculated confinement factors are provided in the Supplementary Figure 2 and Supplementary Table 1, respectively. Fig 8(a) presents the measured maxima of optical gain as a function of current density of the studied LDs. A linear fit is used to extract the differential gain. The change in differential gain with the emission wavelength of the LDs with the MQW design are substantial. The differential gain falls from the value of 10.4 to 2.2 cm/kA when the emission wavelength is changed from blue to cyan. This is in agreement with other reports on the long wavelength LDs [54-57]. One of the major reasons of the drop is the decrease of the <*e1h1*> wavefunction overlap, depicted in Figs 2(a) and 6(a), and consequently the decrease of the IQE shown in Fig 7(c). The differential gain of the LDs incorporating a wide 10.4 nm InGaN QWs are 12.7 and 6.5 cm/kA for the blue and cyan LDs, respectively. The increase in both cases is significant but is larger in the case of the higher indium content QW as predicted in the previous section.

These results, showing a high optical gain, are of vast importance as they indicate that the real-life optoelectronic devices can benefit from the increased IQE of wide InGaN QWs.

## Conclusions

It was shown that, counterintuitively and contrary to the common opinion, wide InGaN QWs have a higher efficiency than their thin counterparts. The increase in the efficiency comes from a transition through excited states with high wavefunction overlap and lower carrier density. It is shown that in a wide QW the carriers occupying the ground states do not recombine but rather screen the built-in electric field. Even after full screening of the field the wavefunction overlap is small enough to prevent recombination through this transition. Instead carriers start to occupy the excited states with a high wavefunction overlap between them. Additionally, the lower carrier density in a wide InGaN QW, compared to normal QWs at the same current densities, ensures a reduction in the nonradiative Auger recombination responsible for the efficiency droop. Furthermore, it was shown that the higher the indium content the higher the wavefunction overlap between the excited states becomes. This result opens a new way to solving the green gap problem in LED efficiency with wide InGaN QWs.

Two sets of experiments have been presented to verify the theoretical predictions. In the first one power dependent photoluminescence was measured for QW structures with various thicknesses. A clear transition from the recombination through the ground states to the excited states was observed with increased thickness of QW proving the predictions of the model. In the second experiment the optical gain of LDs with wide QWs was measured. It was shown that an enhancement of the material gain was achieved for wide QWs emitting in both blue and cyan colors. A more significant increase in the efficiency was observed in the case of higher indium content QWs as predicted by the model.

We hope these findings will initiate interest and give rise to further research on wide InGaN QWs.

## Methods

**Epitaxial growth**

The epitaxial growth was carried out using plasma-assisted molecular beam epitaxy. The MQW samples and blue LDs were grown in a customized VG V90 MBE reactor while the cyan LDs were grown in a customized Veeco Gen20A MBE reactor. Both machines were equipped with two Veeco RF plasma sources. The MQW samples and green LDs were grown on bulk GaN substrates obtained by hydride vapor phase epitaxy while the blue LDs were grown on bulk Ammono-GaN substrates. The InGaN layers were grown at 650˚C in indium-rich conditions whereas GaN and AlGaN layers at 730˚C at gallium-rich conditions. The high quality InGaN growth was carried out by suppling a high active nitrogen flux of up to 2 μm/h in units of equivalent GaN growth rate. Details of InGaN growth mechanism by PAMBE can be found in Ref. [46, 47].

**Photoluminescence measurement**

The PL was measured at room temperature (298 K) with a 325 nm He-Cd laser. The system consisted of a 0.55 m long monochromator coupled with a liquid nitrogen cooled Symphony Si CCD detector. A microscope objective was used to focus the laser light on the sample surface and collect the emitted light. The excitation spot was circular with a diameter of 70 μm.

**Calculations**

Commercially available SiLENSe 5.4 [51] package has been used to calculate the wavefunction overlaps. It calculates the potential shape in an LED at a given voltage by solving the Poisson equation. Afterwards, the Schrödinger equation is solved to calculate the wavefunction distribution and the wavefunction overlap. The default material properties provided with SiLENSE 5.4 had been used. It is important to stress that in our opinion any simulation tool should provide similar results regarding the dynamic screening of polarization charges in wide QWs.

The IQE has been calculated taking into account the ABC model and the change in wavefunction overlap with current density calculated using SiLENSe. The A, B and C parameters have been taken from Ref [53].

**Optical gain measurement**

The optical gain was measured using the Hakki-Paoli technique [59, 60]. The LDs were electrically driven below threshold and the high resolution amplified spontaneous emission (ASE) spectra were collected using a 1 m spectrometer equipped with a 3600 $mm^{-1}$ grating and a CCD camera. An exemplary ASE spectrum of the cyan wide QW LD is presented in Supplementary Figure 3(a). Its magnification showing the adjacent peak in ASE spectrum is presented in Supplementary Figure 3(b). For comparison the lasing spectra collected above threshold is demonstrated in Supplementary Figure 3(c). The gain spectra were calculated based on the ratio of valley to peak from the ASE spectra for each current density and are presented in Supplementary Figure 3(c). Next the maxima of each spectra were extracted and plotted as a function of current density in Figure 8. The internal and mirror losses were subtracted from the values of maximum gain in Figure 8 to clearly show the variation of the differential gain between the LDs.

## Acknowledgements

The authors would like to thank J. Borysiuk for TEM imaging. This work has been partially supported by the Foundation for Polish Science grant TEAM TECH/2016-2/12, the National Centre for Research and Development grants PBS3/A3/23/2015 and LIDER/29/0185/L-7/15/NCBR/2016.

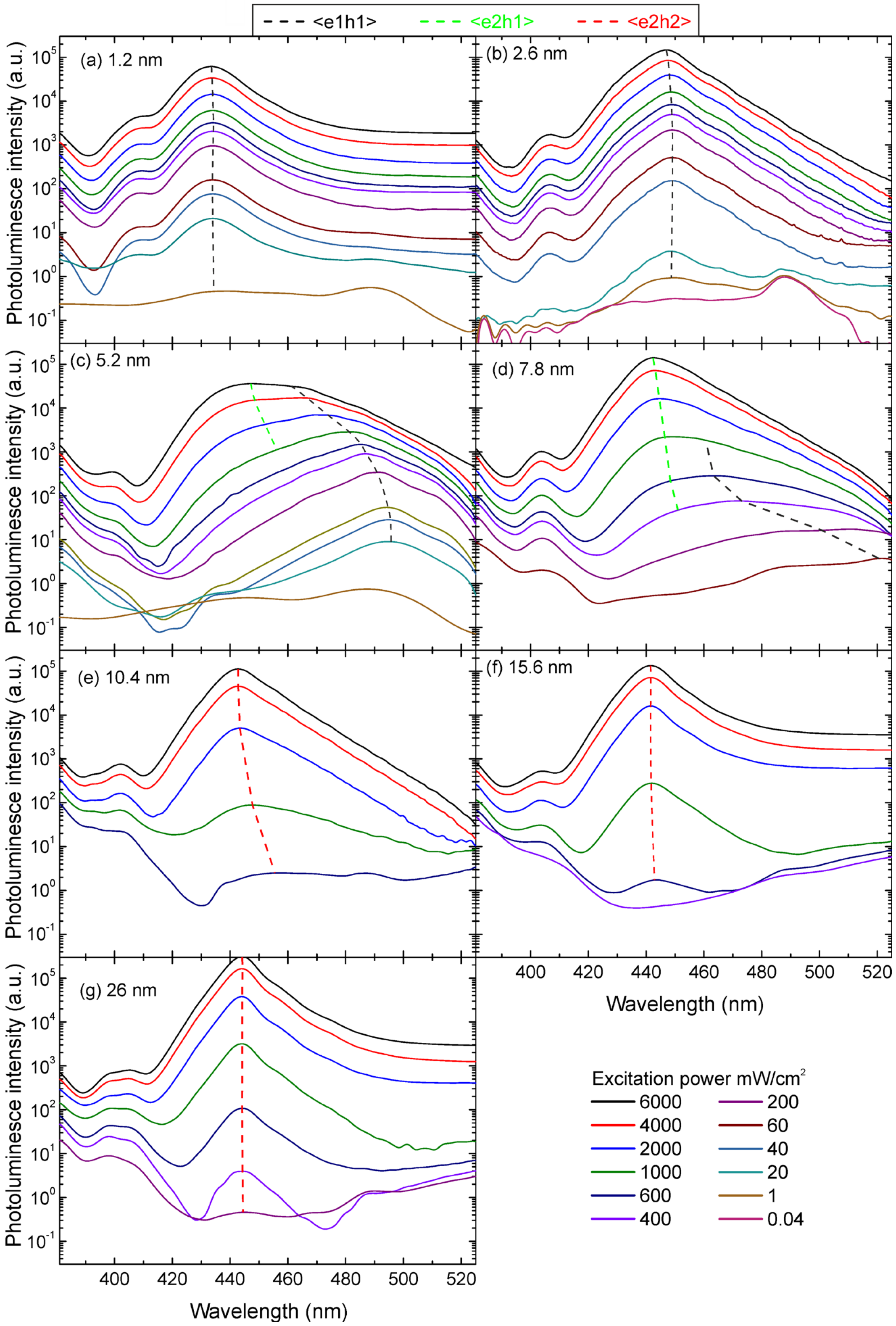

*Supplementary Figure 1. Power dependent photoluminescence for QWs with seven thicknesses: (a) 1.2 nm (b) 2.6 nm, (c) 5.2 nm, (d) 7.8 nm, (e) 10.4 nm, (f) 15.6 nm and (g) 26 nm. The dashed lines depict our interpretation of the observed optical transitions.*

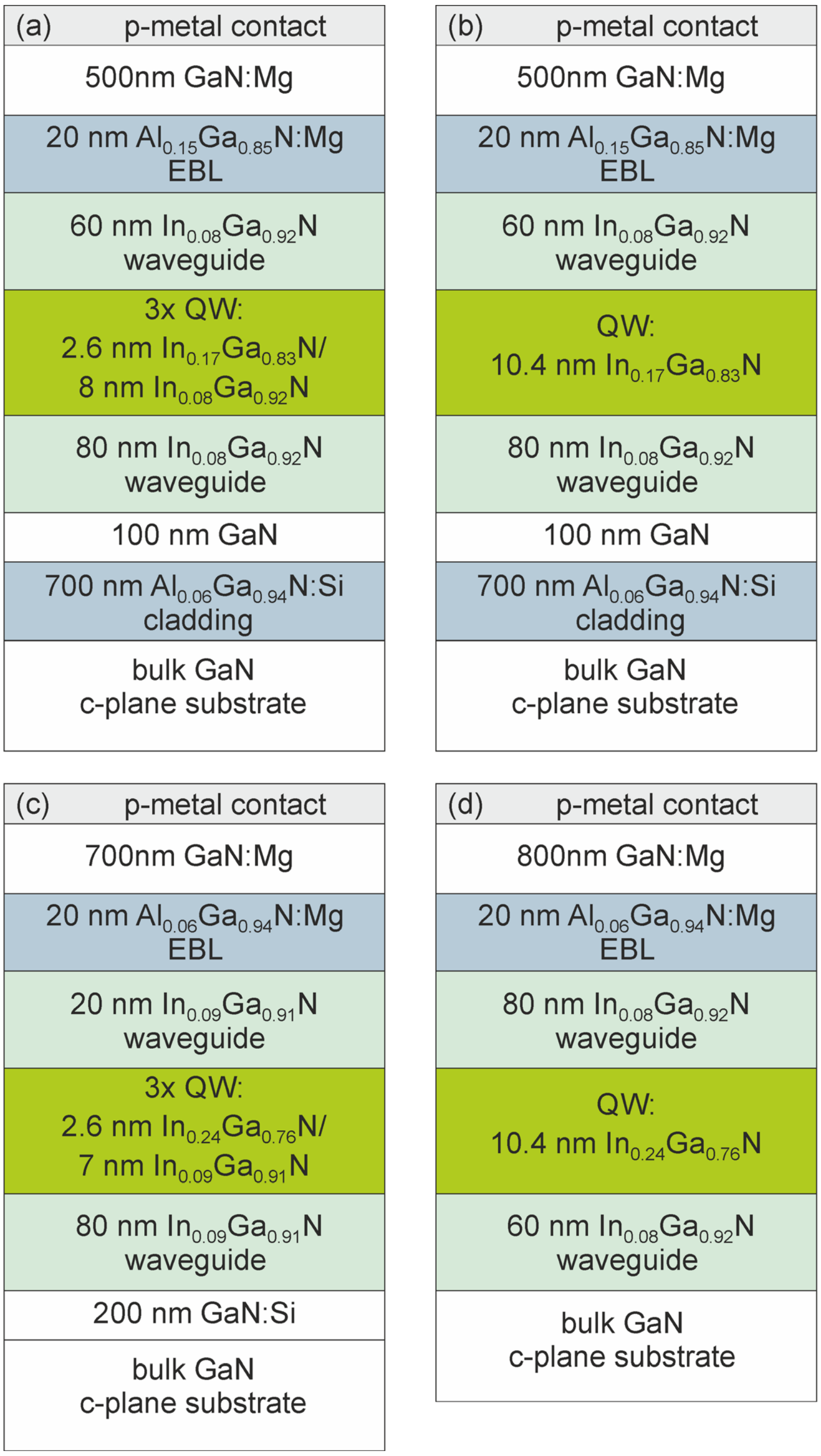

*Supplementary Figure 2. The structure schematics of LDs used for the optical gain experiments. The LDs are identified by their active region in Figure 8 in manuscript: (a) 3x2.6 nm 17% In, (b) 10.4 nm 17% In, (c) 3x3 nm 24% In and (d) 10.4 nm 24% In.*

*Supplementary Table 1. The calculated optical confinement factor of LDs used for the optical gain experiments together with the measured differential gain. The change in the differential gain cannot be justified by the change in confinement factor.*

| LD sample | Calculated optical confinement factor (%) | Measured differential gain (cm/kA) |
|---|---|---|
| 3x2.6nm $In_{0.17}GaN$ | 2.763 | 10.4 |
| 10.4nm $In_{0.17}GaN$ | 3.531 | 12.7 |
| 3x3nm $In_{0.24}GaN$ | 1.914 | 2.2 |
| 10.4nm $In_{0.24}GaN$ | 2.795 | 6.5 |

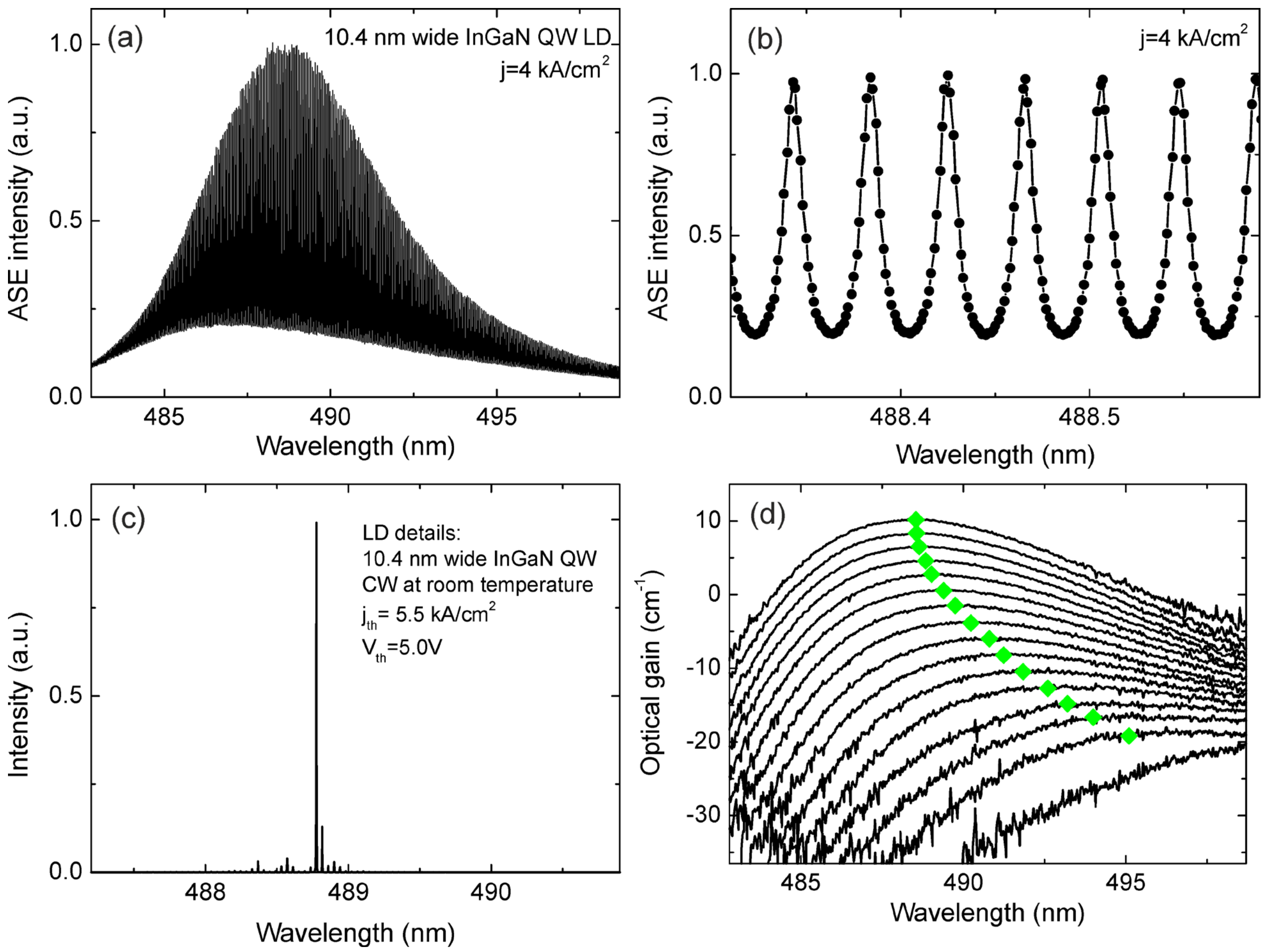

*Supplementary Figure 3. (a) Amplified spontaneous emission (ASE) spectra of the LD with 10.4 nm 24% In quantum well collected below lasing threshold. (b) Magnification of the ASE showing the adjacent peaks used to calculate the optical gain. (c) Lasing spectra collected above lasing threshold. (d) Optical gain spectra collected for current densities ranging from 0.33 to 5.33 in steps of 0.33 kA/cm$^2$. The green diamonds are maximum optical gain used to show the dependence of optical gain on current density in Figure 8.*